\documentclass[aps,reprint,twocolumn,showkeys,superscriptaddress,groupedaddress]{revtex4-1}  % for review and submission
\usepackage{graphicx}  % needed for figures
\usepackage{dcolumn} % needed for some tables
\usepackage{bm}         % for math
\usepackage{amssymb}   % for math
\usepackage{epstopdf}
\usepackage{xcolor}

\begin{document}

\widetext

% the following line is for submission, including submission to the arXiv!!
%\hspace{5.2in} \mbox{Fermilab-Pub-04/xxx-E}

\title{Single-shot Readout of a Superconducting Qubit \\
using a Josephson Parametric Oscillator}

%\title{Josephson parametric oscillator readout of a superconducting qubit\\
%with near single-shot performance without using a quantum-limited amplifier}

\author{Philip Krantz$^{1}$}
\email{philip.krantz@chalmers.se}
\author{Andreas Bengtsson$^{1}$}
\author{Micha\"el Simoen$^{1}$}
\author{Simon Gustavsson$^{2}$}
\author{Vitaly Shumeiko$^{1}$}
\author{W. D. Oliver$^{2,3}$}
\author{C. M. Wilson$^{4}$}
\author{Per Delsing$^{1}$}
\author{Jonas Bylander$^{1}$}
\email{jonas.bylander@chalmers.se}

\address{$^{1}$Microtechnology and Nanoscience, Chalmers University of Technology, Kemiv\"agen 9, SE-41296, Gothenburg, Sweden\\
$^{2}$Research Laboratory of Electronics, Massachusetts Institute of Technology, Cambridge, Massachusetts 02139, USA\\
$^{3}$MIT Lincoln Laboratory, 244 Wood Street, Lexington, Massachusetts 02420, USA\\
$^{4}$Institute of Quantum Computing, University of Waterloo, Waterloo, Ontario N2L 3G1, Canada}       
                             
\date{\today}

\begin{abstract}
\noindent We propose and demonstrate a new read-out technique for a superconducting qubit by dispersively coupling it to a Josephson parametric oscillator. 
We employ a tunable quarter-wavelength superconducting resonator and modulate its resonant frequency at twice its value with an amplitude surpassing the threshold for parametric instability.
We map the qubit states onto two distinct states of classical parametric oscillation: 
one oscillating state, with $185\pm15$ photons in the resonator, and one with zero oscillation amplitude.
This high contrast obviates a following quantum-limited amplifier.
We demonstrate proof-of-principle, single-shot readout performance, and present an error budget indicating that this method can surpass the fidelity threshold required for quantum computing.
\end{abstract}

%\pacs{42.50.Lc, 42.50.Pq, 42.65.Yj, 03.67.Lx}
\maketitle

The readout scheme for quantum bits of information (qubits) constitutes one essential component of a quantum-information processor~\cite{Divincenzo2000}.
During the course of a quantum algorithm, qubit-state errors need to be corrected; 
%in many implementations this is done by the frequent application of coherent \textcolor{red}{quantum-feedback operations, where the quantum state is being used to evolve the system towards a desired target state~\cite{Lloyd2000}.}
in many implementations this is done by quantum error correction, where each operation is based on the outcomes of stabilizer measurements that indicate the qubit errors. The stabilizers must therefore be determined in a ``single shot'' -- without averaging of the output signals of repeated measurements on identically prepared qubits -- with fidelity exceeding approximately 99\,\%~\cite{Kelly2015}.\\
\indent The commonly used measurement scheme for a superconducting qubit coupled to a linear microwave resonator does not, by itself, offer single-shot measurement performance.
The qubit imparts a state-dependent (``dispersive") frequency shift on the resonator, which can be determined by applying a probe signal and measuring the reflected or transmitted signal, although only for weak probing, rendering an inadequate signal-to-noise ratio (SNR)~\cite{Gambetta2006, Boissonneault2009}.\\
\indent Researchers have addressed the problem of insufficient SNR in essentially two ways.
One approach is to feed the weak output signal into a following,  parametric linear amplifier that adds only the minimum amount of noise allowed by quantum mechanics~\cite{Vijay2011, Lin2013, Lin2014, Bergeal2010}.
Another approach is to insert a nonlinear element into the system and apply a strong drive tone, such that the resonator enters a bistable regime, hence enhancing the detection contrast~\cite{Lupascu2007, Mallet2009, Murch2012, Reed2010, Boissonneault2010}.\\ 
\indent In this paper we propose and demonstrate a simplified readout technique in which a superconducting qubit is directly integrated into a Josephson parametric oscillator (JPO).
We map the qubit states onto the ground and excited states of the oscillator, and demonstrate proof-of-concept, single-shot readout performance (SNR $>1$).
%We obtain 81.5$\,\%$ qubit-state discrimination, however, from the error analysis we infer a read-out fidelity of $98.7\pm1.2\,\%$, taking into account known and reparable errors due to qubit initialisation and decoherence ($17.2\pm1.2\,\%$). The remaining errors, which are due to switching events in the oscillator ($1.2\pm0.3\,\%$), can be eliminated by improving the data-aquisition protocol --- see the Discussion and Supplementary Information. \textcolor{red}{ Given a realistically achievable qubit relaxation time, $T_1 = 50\,\mu$s, this would bring the read-out fidelity to $99.4\,\%$.}\\
We obtain 81.5$\,\%$ qubit-state discrimination for a read-out time $\tau = 600\,$ns; 
however, from the error analysis we infer a read-out fidelity of $98.7\pm1.2\%$, taking into account known and reparable errors due to qubit initialisation and decoherence ($17.2\pm1.2\,\%$). 
A realistically achievable qubit-relaxation time, $T_1 = 50\,\mu$s, and a Purcell bandpass filter would reduce these errors from $17.2\,\%$ to $< 0.5\,\%$, as well as shorten the required read-out time to $\tau < 100\,$ns.
The remaining errors, which are due to switching events in the oscillator ($1.2\pm0.3\,\%$), can be eliminated by improving the data-aquisition protocol - see Discussion and Supplementary Information.
These qubit and detection improvements would bring the read-out fidelity to $\approx 99.5\,\%$.\\
\indent Our readout scheme relies on parametric pumping of a frequency-tunable resonator by modulation of its inductance.
The pumping amplitude exceeds the threshold for parametric instability, the point above which the resonator oscillates spontaneously, even in the absence of an input probe signal.
This instability threshold is controlled by the state of the qubit, whose ground and excited states correspond to the nonoscillating and oscillating states of the resonator, respectively. 
In our measurement, the oscillating state produces a steady-state resonator field corresponding to $185\pm15$ photons, whose output we can clearly distinguish from the nonoscillating state when followed by a commercial semiconductor amplifier, eliminating the need for a quantum-limited amplifier.
Conceptually, this method can yield arbitrarily large contrast due to the parametric instability, and moreover, only requires a pump but no input signal.\\
%\textcolor{red}{Conceptually, this method can yield arbitrarily large contrast due to the parametric instability, ease the need for a high quantum efficiency on the following measurement chain, and only requires a pump but no input signal.}\\
%
%\indent This readout scheme is well aligned with scalable, multi-qubit implementations. Parametric oscillators can be readily frequency-multiplexed~\cite{Schmitt2014} and allow for a simplified experimental setup (compared to conventional microwave reflectometry) without a separate input port to the resonator, bulky microwave circulators \textcolor{red}{(except for isolators)}, or a following parametric amplifier. 
%It is also possible to manipulate the qubit via the flux-pumping line, which further reduces the required number of cables and interconnects.
%
\indent This readout scheme is well aligned with scalable, multi-qubit implementations. Parametric oscillators can be readily frequency-multiplexed~\cite{Schmitt2014} and allow for a simplified experimental setup (compared to conventional microwave reflectometry) without a separate input port to the resonator or a following parametric amplifier, and consequently also without additional bulky microwave circulators that would normally route the input and parametric-pumping tones. It is also possible to manipulate the qubit via the flux-pumping line only, which further reduces the number of cables and interconnects.

\begin{figure}[t!]
\includegraphics[width=1\linewidth]{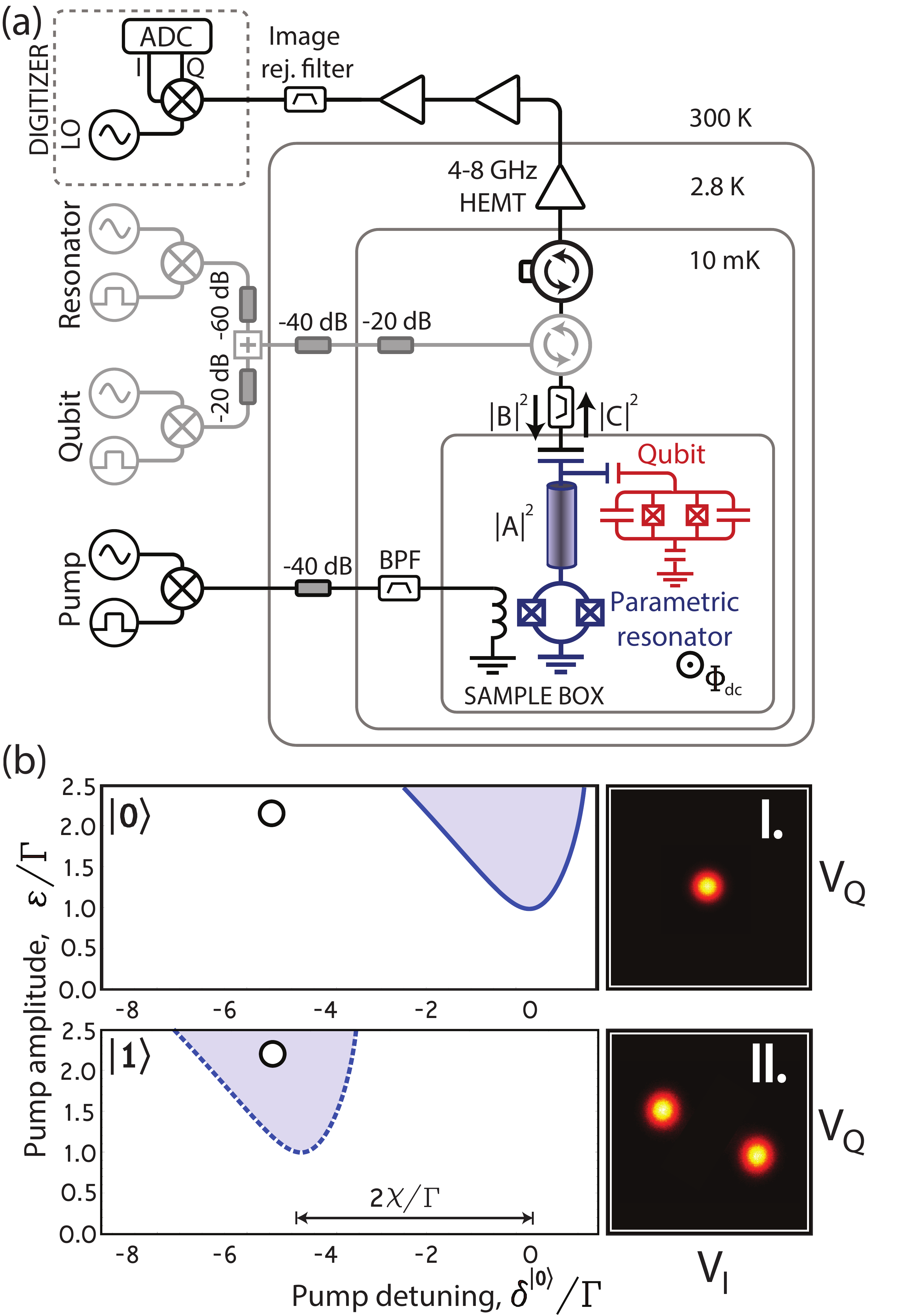} 
\caption{\label{fig:SetupAndReadoutPrinciple} Experimental setup and readout mechanism. 
\textbf{(a)} Schematic of the cryogenic microwave reflectometry setup. The transmon qubit (red) is capacitively coupled to the coplanar waveguide parametric resonator (blue). The input and output flows of photons are denoted $\left|B\right|^2$ and $\left|C\right|^2$, respectively, whereas the number of photons in the resonator is denoted $\left|A\right|^2$. 
The output signal is acquired using heterodyne detection of the amplified microwave signal. 
The components drawn in lighter gray are those that are rendered unnecessary by the JPO readout method, thereby offering a simplified experimental setup (see text).
\textbf{(b)} Parametric-oscillation regions for the qubit ground state $\left|0\rangle\right.$ (solid blue line) and excited state $\left|1\rangle\right.$ (dashed blue line), respectively. These blue lines represent the instability boundaries, $\epsilon = \epsilon_{\mbox{\tiny{th}}}$, where the number of steady-state solutions to Eq. (\ref{eq:Eq1}) changes --- see Eq. (\ref{eq:ParametricInstabilityThreshold}) in Methods. The two panels on the right are measured [I,Q]-quadrature voltage histograms of the device output for the pump bias point indicated by the circles, revealing two different oscillator states: 
\textbf{I.} Outside of the region of parametric oscillations, the resonator is ``quiet" ($|A|^2 = 0$). 
\textbf{II.} Within the region, the resonator has two oscillating states ($|A|^2 > 0$), with a phase difference of $\pi$ radians --- see further Fig.~\ref{fig:Histograms}.}
\end{figure}

\section*{\label{sec:Results}Results}
\textbf{The Josephson parametric oscillator (JPO).}
Our device consists of a quarter-wavelength ($\lambda/4$), superconducting coplanar waveguide resonator, shorted to ground in one end via two parallel Josephson tunnel junctions (JJs) --- see Fig.~$\ref{fig:SetupAndReadoutPrinciple}$(a).  The JJs form a superconducting quantum interference device (SQUID), which acts as a variable Josephson inductance, $L_J (\Phi,I(\phi)) = \Phi_0 / \left[ 2\pi \cos (\pi\Phi/\Phi_0) \sqrt{ I_0^2 - I^2(\phi)}\right]$, where $I_0$ is the critical current and $\Phi_0$ is the flux quantum. 
This inductance can be controlled by the external magnetic flux through the SQUID loop, $\Phi(t) = \Phi_{\mbox{\tiny{dc}}} + \Phi_{\mbox{\tiny{ac}}}(t)$, and by the superconducting phase difference across the JJs, $\phi(t)$, via its current--phase relation, $I(t) = I_0 \sin\phi(t)$.

Time-varying modulations of $\Phi$ and $\phi$   -- ``parametric pumping" --  affect the resonator dynamics, albeit in rather different ways; moreover, the Josephson inductance is indeed both \emph{parametric} and \emph{nonlinear}. We explain these differences in the Discussion section below.
The resonant frequency of the JPO is parametrically modulated via the magnetic flux, $\Phi(t)$, which can lead to frequency mixing as well as parametric effects such as noiseless amplification of a signal, frequency conversion, and instabilities~\cite{Sandberg2008,Palacios-Laloy2008, Wilson2010, Wilson2011, Lin2013, Krantz2013}.

\indent The state of the JPO has a rich dependence on several parameters, some of which was studied recently, both theoretically~\cite{Dykman1998, Wustmann2013} and experimentally~\cite{Wilson2010, Krantz2013,Lin2014}.
The equation of motion for the intra-resonator electric field amplitude, $A$,
can be written as
\begin{equation} \label{eq:WS_equation}
i\dot{A} + \epsilon A^{*} + \delta A + \alpha \left| A\right|^2 A + i\Gamma A = \sqrt{2 \Gamma_0}B(t).
\label{eq:Eq1}
\end{equation}

\noindent Here $\epsilon$ is proportional to the externally applied pump amplitude, $\Phi_{\mbox{\tiny{ac}}}$, which modulates the resonant frequency parametrically at close to twice its value, $\omega_p \approx 2 \omega_r$ (degenerate pumping), and $\delta = \omega_p/2 - \omega_r$ is the resonator's detuning from half of the pump frequency. 
The field amplitude, $A$, and its complex conjugate, $A^{*}$, are slow variables in a frame rotating at $\omega_p/2$, and $|A|^2$ is the equivalent number of photons in the resonator.
The Duffing parameter, $\alpha$, associated with a cubic field nonlinearity, arises from the nonlinear Josephson inductance.
The linear damping rate has two components, $\Gamma = \Gamma_0 + \Gamma_R$, where $\Gamma_0/2\pi = 1.02\,$MHz is the external damping rate, associated with the photon decay through the coupling capacitor, and $\Gamma_R/2\pi = 0.30\,$MHz is the internal loss rate. 
The equation's right-hand side represents the input probe signal, such that $\left|B(t)\right|^2$ has units of photons per second. 
The output flow of photons per second, $|C(t)|^2$, is given by $C(t) = B(t) - i\sqrt{2\Gamma_0}A$.\\
\indent For low pumping amplitude, below the parametric instability threshold, $\epsilon < \epsilon_{\mbox{\tiny{th}}}$,  this device works as a phase-sensitive parametric amplifier (JPA) for an input $B(t)$ at signal frequency $\omega_s = \omega_p/2$~\cite{Yamamoto2008, Sandberg2008, Palacios-Laloy2008, Wilson2010, Lin2013}.
Note, however, that we keep $B(t) \! = \! 0$ in the measurements reported here.
For a pumping amplitude exceeding the threshold, $\epsilon > \epsilon_{\mbox{\tiny{th}}}$, spontaneous parametric oscillations set in --- see Fig.~\ref{fig:SetupAndReadoutPrinciple}(b) and Eq. (\ref{eq:ParametricInstabilityThreshold}) in Methods.
The resonator field builds up exponentially in time, even in the absence of an input probe signal until it becomes limited by the Duffing and pump-induced nonlinearities and reaches a steady state~\cite{Wilson2010,Krantz2013}.\\
\indent We connected a transmon qubit capacitively to the resonator \cite{Koch2007} --- see Fig.~\ref{fig:SetupAndReadoutPrinciple}(a). The state of the JPO (oscillating or nonoscillating) can then be controlled by the qubit-state-dependent, dispersive frequency shift, $\chi$, which the qubit exerts on the resonator~\cite{Blais2004,Wallraff2004}. 
When the JPO is being pumped above the threshold for parametric oscillation, with amplitude $\epsilon$ and frequency detuning $\delta$, then a change of qubit state effectively pulls the resonator to a different value of the detuning, outside of the region of parametric oscillations --- see Fig.~\ref{fig:SetupAndReadoutPrinciple}(b). 
We denote the qubit-state dependent detunings by $\delta^{|0\rangle} = \delta - \chi$ and $\delta^{|1\rangle} = \delta + \chi$. 
The resulting mapping of the qubit state onto the average number of photons in the resonator provides us with a novel qubit-state readout mechanism, which we exploit in this work.\\

\begin{figure}
\includegraphics[width=1\linewidth]{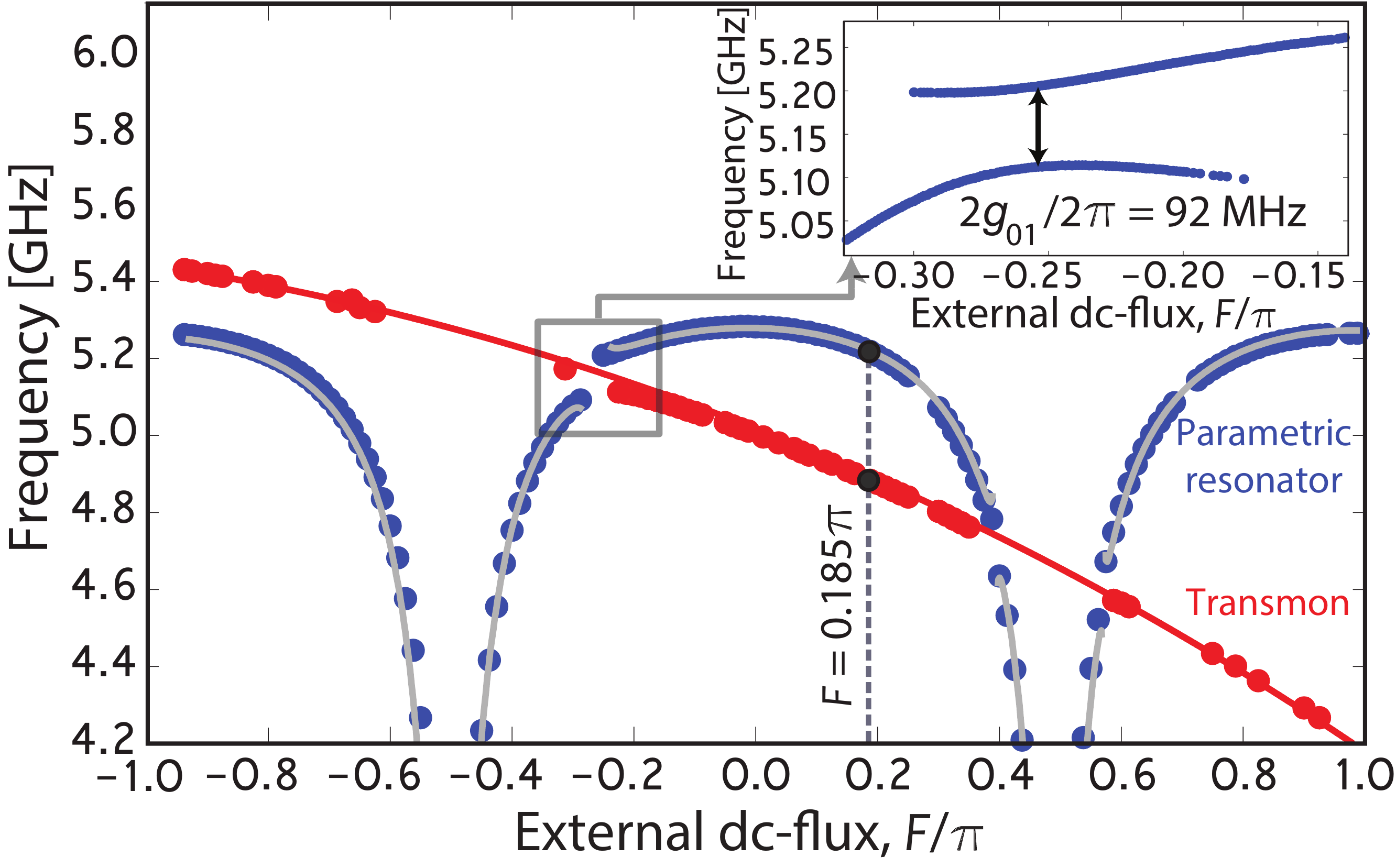} 
\caption{\label{fig:SpectroscopyAndAvoidedCrossing} Combined frequency spectrum obtained from qubit spectroscopy of the transmon qubit (in red) and through standard reflectometry of the resonator (in blue). The solid red and grey lines are fits. The dashed grey line, at resonator flux bias $F = 0.185\pi$, indicates the bias point at which we later demonstrate the readout method. \textbf{Inset:} Vacuum-Rabi splitting around the flux-bias point where the transmon frequency crosses that of the resonator. The minimum frequency splitting yields a qubit-resonator coupling $g_{01}/2\pi = 46\,$MHz.}
\end{figure}

\textbf{Characterisation of qubit and JPO.} 
\indent The device and cryogenic experimental setup are depicted in Fig.~\ref{fig:SetupAndReadoutPrinciple}(a).
The sample is thermally anchored to the mixing chamber of a dilution refrigerator with a base temperature of 10$\,$mK. The parametric $\lambda/4$ resonator (in blue) is capacitively coupled to the transmission line ($C_c = 11.9$~fF), yielding an external quality factor $Q_{\mbox{\tiny{ext}}} = \omega_r/2\Gamma_0 = 2\,555$. A transmon qubit (in red) is also coupled near this end of the resonator.\\
\indent The resonator output signal is amplified using a 4--8 GHz high-electron-mobility transistor amplifier, with a noise temperature $T_N = 2.2\,$K, followed by two room-temperature amplifiers. 
We detect the outgoing signal using heterodyne mixing.
The signal is first downconverted to a frequency $\left(\omega_{\mbox{\tiny{RF}}} - \omega_{\mbox{\tiny{LO}}}\right)/2\pi = 187.5\,$MHz; then the [I,Q]-quadrature voltages are sampled at 250$\,$MS$\,$s$^{-1}$, before they are digitally downsampled at a rate of 20$\,$MS$\,$s$^{-1}$.\\
\indent We first characterise the transmon spectroscopically --- see Fig.~\ref{fig:SpectroscopyAndAvoidedCrossing} --- from which we extract the Josephson and charging energies, $E_{J}/2\pi$ = 9.82$\,$GHz and $E_{C}/2\pi$ = 453$\,$MHz, respectively. From the vacuum-Rabi splitting, we extract a qubit$\--$resonator coupling rate $g_{01}/2\pi$ = 46$\,$MHz --- see inset in Fig. \ref{fig:SpectroscopyAndAvoidedCrossing}.\\
\indent Next, we fit the frequency tuning curve of the resonator (with the qubit in the $\left|0\rangle\right.$-state) to the relation
\begin{equation}
\omega_{r}^{|0\rangle}(F) = \omega_r(F) - g_{01}^2 / \Delta(F) ,
\label{eq:ResonatorTuningCurve}
\end{equation}

\noindent where $F = \pi \Phi_{\mbox{\tiny{dc}}}/\Phi_0$ denotes the static flux bias, normalised to the magnetic flux quantum. The effective dispersive shift due to the qubit is
\begin{equation}
\chi(F) = -\frac{g_{01}^2}{\Delta(F)}\left(\frac{E_{C}}{\Delta(F) - E_{C}}\right),
\label{eq:DispersiveShift}
\end{equation}

\noindent which, in turn, depends on the qubit--resonator detuning, $\Delta(F) = \omega_a(F') - \omega_r(F)$, with $F' = F/8.88 + 0.58$ representing the effective magnetic flux of the transmon. Moreover, the qubit and resonator frequency spectra are well approximated by~\cite{Koch2007,Wallquist2006}
\begin{equation}
\omega_a(F') \approx \sqrt{8E_{J}\left|\cos(F')\right|E_{C}} - E_{C},
\label{eq:QubitTuningCurve}
\end{equation}
\begin{equation}
\omega_r(F) \approx \frac{\omega_{\lambda/4}}{1 + \gamma_0/\left|\cos(F) \right|},
\label{eq:ResonatorTuningCurve2}
\end{equation}

\begin{figure}
\includegraphics[width=1\linewidth]{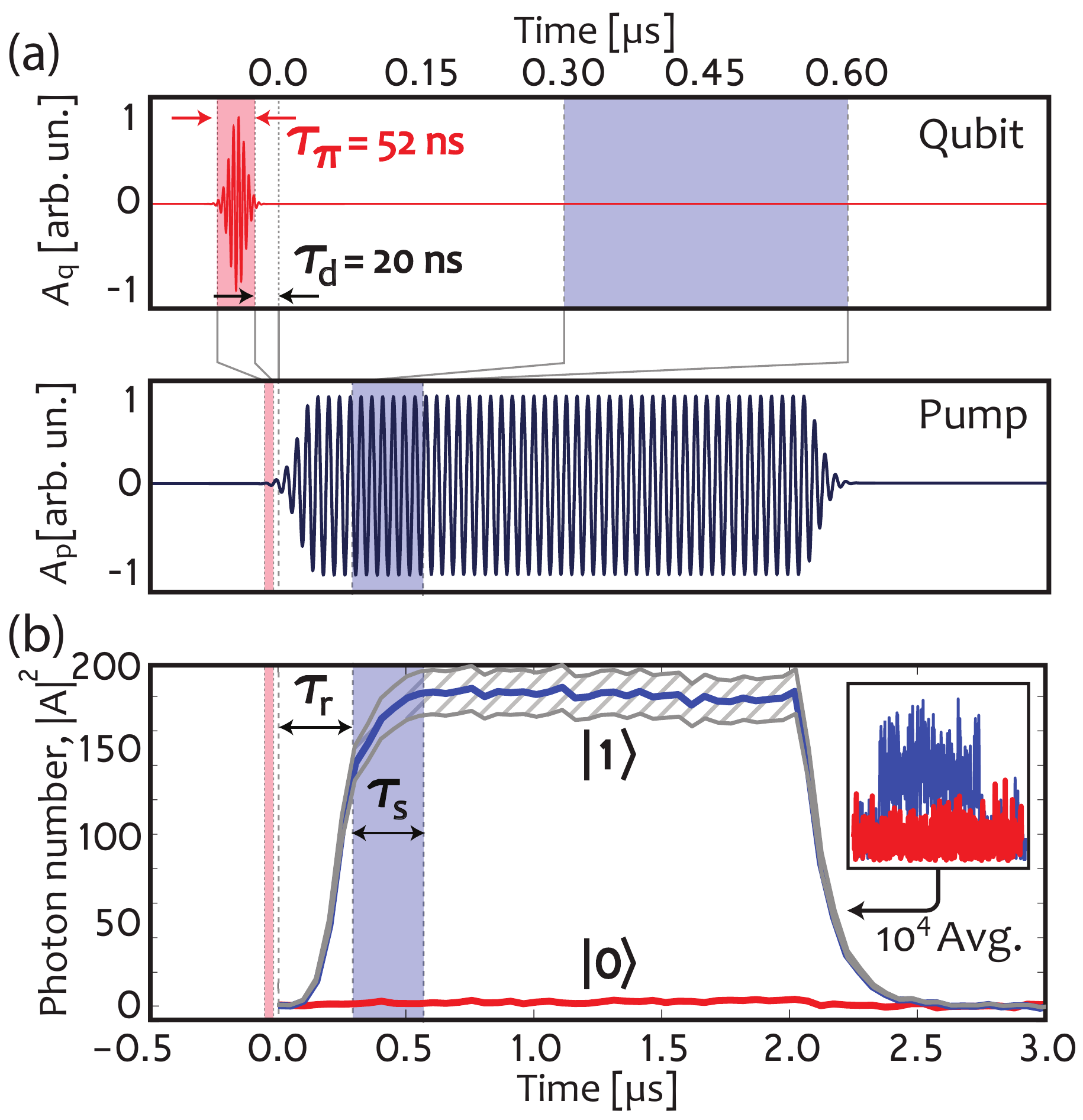}
\caption{\label{fig:PulseSequences}   Qubit readout by the Josephson parametric oscillator. \textbf{(a)} Pulse sequence: The qubit $\pi$-pulse (in red), with Gaussian edges and a plateau of duration $\tau_{\pi} = 52\,$ns, is followed by a short delay, $\tau_{d} = 20\,$ns, before the pump is turned on at time $t = 0$. \textbf{(b)} The solid blue and red traces show the inferred photon number, $|A|^2$, in the resonator, with and without a prior $\pi$-pulse on the qubit, respectively. Note that the resonator latches, once it has entered into the oscillating state, and remains there even if the qubit relaxes. The traces are the result of $10^4$ averages of the raw data; the inset shows a single instance of the raw data on the same time axis as the main plot. Prior to the sampling window of width $\tau_{s} = 300\,$ns, a delay $\tau_{r} = 300\,$ns is added to avoid recording the transient oscillator response. The hatched region around the average photon number represents our uncertainty, originating from the amplifier gain calibration --- see Methods and Supplementary Fig. 3.}
\end{figure}

\noindent where $\omega_{\lambda/4}/2\pi = 5.55\,$GHz is the bare resonant frequency (in absence of the SQUID), and $\gamma_0 = L_{J}(F\!\!=\!\!0)/L_{\mbox{\tiny{r}}} = 5.3\pm0.1\,\%$ is the inductive participation ratio between the SQUID (at zero flux) and the resonator. The solid grey and red lines in Fig.~\ref{fig:SpectroscopyAndAvoidedCrossing} are fits to Eqs.~(\ref{eq:ResonatorTuningCurve}) and (\ref{eq:QubitTuningCurve}), respectively.\\

\begin{figure*}
\includegraphics[width=1\linewidth]{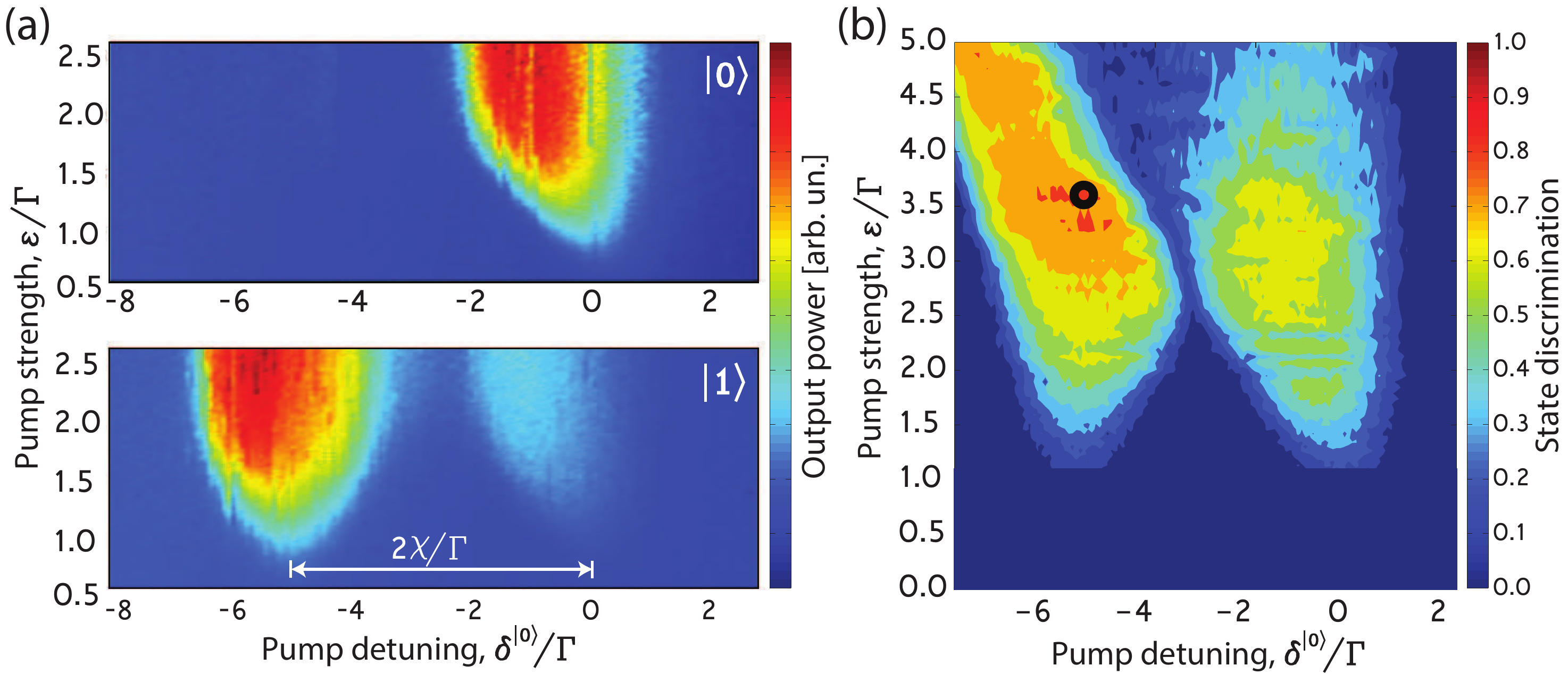}
\caption{\label{fig:ParametricRegions} Parametric oscillations, and state discrimination. \textbf{(a)} Output field of the resonator when the qubit is in its ground state $\left|0\rangle\right.$ (top panel) and excited state $\left|1\rangle\right.$ (bottom panel). \textbf{(b)} Contour plot of the state discrimination within the two parametric oscillation regions. The black circle in the left region, located at $\delta^{\left|0\rangle\right.}/\Gamma = -5.34, \epsilon/\Gamma = 3.56$, represents the bias point used throughout the rest of the analysis and in Fig. \ref{fig:PulseSequences}(b). The state discrimination in this point is 81.5$\%$.}
\end{figure*}

\begin{figure}[t!]
\includegraphics[width=1\linewidth]{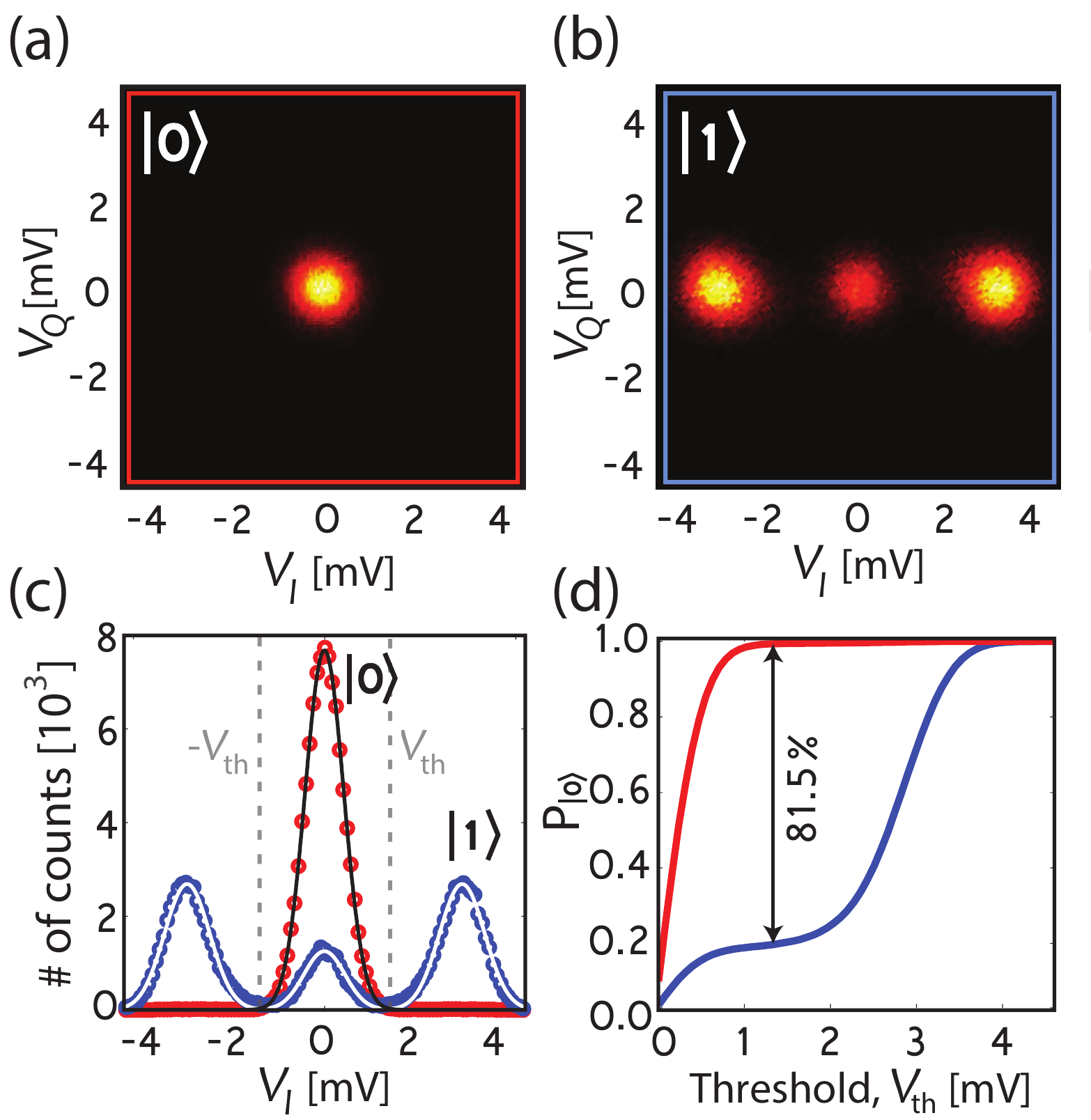}
\caption{\label{fig:Histograms} Quadrature voltage histograms of the parametric oscillator output, collected after digital sampling. The pump bias point was $\delta^{|0\rangle}/\Gamma = -5.34, \, \epsilon/\Gamma = 3.56$ --- see Fig. \ref{fig:ParametricRegions}(b). In panel \textbf{(a)}, the qubit was in its ground state; in \textbf{(b)}, a $\pi$ pulse was applied prior to the readout pulse. \textbf{(c)} 1D-histograms of the in-phase voltage component, $V_{I}$, from the quadrature histograms in (a) and (b). The black and white solid lines are Gaussian fits, from which we extracted a signal-to-noise ratio of 3.39. \textbf{(d)} Cumulative distribution functions, corresponding to the $|0\rangle$ and $|1\rangle$ states, obtained by sweeping a threshold voltage, $V_{\mbox{\tiny{th}}}$, from the center of the two histograms ($V_{I} = 0$). The maximum separation between the two S-curves yields a state discrimination of 81.5$\,\%$.}
\end{figure}

\textbf{Single-shot qubit readout.} We now demonstrate our method for reading out the qubit with the JPO. We choose a static-flux bias point $F = 0.185\,\pi$ for the resonator SQUID, corresponding to a resonant frequency $\omega_{r}^{|0\rangle}/2\pi = 5.218\,$GHz and qubit transition frequency $\omega_a/2\pi = 4.885\,$GHz --- see dashed grey line in Fig.~\ref{fig:SetupAndReadoutPrinciple}(a). Consequently, the qubit--resonator detuning is $\Delta/2\pi = -334\,$MHz, and the effective dispersive shift is $2\chi/2\pi = -7.258\,$MHz. We measured a Purcell-limited qubit relaxation time, $T_1 = 4.24\pm0.21\,\mu$s, and Ramsey free-induction decay time $T_{2}^{*} = 1.66\pm0.32\,\mu$s --- see Methods, Supplementary Fig. 5, and Table 3.\\
\indent To operate the parametric oscillator as a high-fidelity qubit-readout device, we must be able to map the states of the qubit onto different states of the oscillator, which we must then clearly distinguish. We encode the qubit ground state $|0\rangle$ in the ``quiet" state (the empty resonator) and the excited state $|1\rangle$ in the ``populated" state of the resonator. Figure \ref{fig:PulseSequences}(a) shows the pulse sequence for qubit manipulation and readout, and Fig.~\ref{fig:PulseSequences}(b) shows the resulting output from the JPO, operated with the pump settings $\delta^{|0\rangle}/\Gamma = -5.34, \, \epsilon/\Gamma = 3.56$.\\
\indent The populated oscillator in Fig.~\ref{fig:PulseSequences}(b) contains 185$\pm15$ photons. We obtained this estimate from a comparison between the probe-amplitude dependence of the resonant frequency and the expected photon-number dependence of the Duffing shift --- see Methods and Supplementary Fig. 3. This number of photons should be compared to $|A|^2$ = 200$\pm3$ photons, which is the solution to Eq. (\ref{eq:Eq1}) in the steady state ($\dot{A}=0$).\\
\indent In order to achieve such clear qubit-state discrimination as in Fig.~\ref{fig:PulseSequences}(b), we needed to make a judicious choice of flux-bias point, $F$, to mitigate the effects of two nonlinear shifts of the resonant frequency~\cite{Krantz2013}. The Duffing shift dominates when $F \rightarrow \pm\pi/2$, whereas a pump-induced frequency shift dominates when $F \rightarrow 0$. These shifts can move the resonator away from the proper pump condition, thereby effectively restricting the output power --- see Methods and Supplementary Fig. 2.\\
\indent Moreover, the qubit$\--$resonator detuning should be in the dispersive regime ($\Delta \gg g_{01}$), in which the qubit state controls the resonant frequency of the resonator.
Yet it must yield a sufficiently large dispersive shift, $\chi > \Gamma$ (Eq. \ref{eq:DispersiveShift}), to produce clearly distinguishable output levels, corresponding to the $|0\rangle$ and $|1\rangle$ states.
For our chosen flux-bias point, we identify the optimal pump settings by mapping out the parametric oscillation region as a function of pump frequency and amplitude --- see Fig.~\ref{fig:ParametricRegions}(a).\\
\indent An interesting feature is present within the left half of Fig.~\ref{fig:ParametricRegions}(a)  (where the populated resonator encodes $| 1\rangle$).
Here, when the qubit is initially in the $| 1\rangle$ state, the resonator latches into its oscillating state for as long as the pump is kept on, and does not transition into its quiet state when the qubit relaxes, as one might have expected.
This latching is shown by the blue trace in Fig.~\ref{fig:PulseSequences}(b). We attribute it to the existence of a tri-stable oscillation state \cite{Wustmann2013, Wilson2010}, associated with red detuning of the above-threshold region for the $|0\rangle$ state. 
When the qubit relaxes, there occurs an instantaneous shift of the pseudopotential for the amplitude $A$, from bi-stable (with two $\pi$-shifted, finite-amplitude states; see Fig.~\ref{fig:SetupAndReadoutPrinciple}(b), panel II.) to tri-stable (with one additional zero-amplitude state).
The field's initial condition at the time of this shift, $A \neq 0$, causes the resonator to maintain its oscillating state. 
A separate study of this latching feature will be reported elsewhere.
\indent We evaluate the obtainable state discrimination by collecting quadrature-voltage histograms at every point within the two regions of parametric oscillations in the [$\delta, \, \epsilon$]-plane --- see Fig.~\ref{fig:ParametricRegions}(b). 
We choose the pump operation point $\delta^{|0\rangle}/\Gamma = -5.34, \, \epsilon/\Gamma = 3.56$, indicated by the black circle, and show the characterization in detail in Fig.~\ref{fig:Histograms}. 
In this point, the state discrimination has reached a plateau around 81.5$\%$.
Each histogram in Fig.~\ref{fig:Histograms}(a--b) contains in-phase ($V_{I}$) and quadrature ($V_{Q}$) voltage measurements from $10^5$ readout cycles, with each measurement being the mean quadrature voltage within the sampling time $\tau_s$ (blue window in Fig.~\ref{fig:PulseSequences}). 
We project each of the 2D-histograms onto its real axis, and thus construct 1D-histograms of the $V_I$ component --- see Fig.~\ref{fig:Histograms}(c).
We can then extract a signal-to-noise ratio, SNR = $|\mu_{|1\rangle} - \mu_{|0\rangle}|/( \sigma_{|1\rangle} + \sigma_{|0\rangle})$ = 3.39, where $\mu$ and $\sigma$ denote the mean value and standard deviation, respectively, of the Gaussians used to fit the histograms. The peak separation of the histograms gives a confidence level of 99.998$\%$ for the readout fidelity. The peak appearing in the center of the blue trace arises mainly from qubit relaxation prior to and during the readout. 
We analyze this and other contributions in the next section, as well as in Supplementary Note 1 and Fig. 4.\\
\indent To extract the measurement fidelity from the histograms, we plot the cumulative distribution function of each of the two traces in Fig.~\ref{fig:Histograms}(c), by summing up the histogram counts symmetrically from the center and outward, using a voltage threshold, $V_{\mbox{\tiny{th}}}$. From these sums, we obtain the S-curves of the probability to find the qubit in its ground state as a function of the voltage threshold value  --- see Fig.~\ref{fig:Histograms}(d). 
We define the fidelity of the measurement as the maximum separation between the two S-curves.\\

\section*{Discussion}
%\textbf{Analysis of readout fidelity.} 
To evaluate the fidelity of the readout itself, as compared to the fidelity loss associated with qubit errors, we now present an error budget. From the histograms in Fig.~$\ref{fig:Histograms}$(c), we can account for 81.5$\,\%$ of the population, thus missing 18.5$\,\%$. To understand the remaining contributions, we run a Monte Carlo simulation of the qubit population, consisting of the same number of $10^5$ readout cycles as in the measured histograms. The simulation results are binned in the same way as the measurements, using the Gaussian fits as boundaries, and taking into account the following statistics: (i) qubit relaxation and preparation errors, (ii) thermal population of the qubit, (iii) spurious switching events by $\pi$ radians of the oscillator phase during readout (yielding a reduced sampled voltage), and (iv) peak-separation error due to the limited signal-to-noise ratio.\\
\indent We find that the main contribution to the loss of fidelity is due to qubit relaxation prior to and during the readout. From the measured relaxation time, $T_1 = 4.24\pm0.21\,\mu$s, we obtain a fidelity loss of $11.6\pm0.5\,\%$. 
However, this error can be reduced substantially (to $<\!0.5\,\%$) by introducing a Purcell bandpass filter~\cite{Reed2010a, Jeffrey2014, Sete2015} at the output of the JPO; since the qubit is detuned from the JPO, this decreases its relaxation into the 50-$\Omega$ transmission line. Such a filter would allow us to increase the resonator damping rate, $\Gamma_0$, substantially reducing the readout time without compromising $T_1$. This is shown in Supplementary Note 2 and Table 4.
%Note, however, that an increased resonator damping rate yields an increased width of the parametric oscillation region: consequently, the qubit--resonator coupling, $g_{01}$, needs to be increased accordingly to result in a sufficiently large dispersive frequency shift. \\ 
Note, however, that an increased resonator damping rate yields an increased width of the parametric oscillation region: consequently, the qubit--resonator coupling, $g_{01}$, and detuning, $\Delta$, need to be chosen accordingly to render a sufficiently large dispersive frequency shift. \\ 
\indent From the simulation, we further attribute $4.5\pm0.3\,\%$ to qubit preparation errors. Another 1.1$\pm0.4\,\%$ can be explained from thermal population of the qubit; the effective qubit  temperature is $T_q = 45\pm3\,$mK. By adding these fidelity loss contributions due to the qubit to the measured state discrimination, we can account for 81.5$\,\%$ + 11.6$\pm0.5\,\%$ + 4.5$\pm0.3\,\%$ + 1.1$\pm0.4\,\%$ = 98.7$\pm1.2\,\%$.\\
\indent There are also errors introduced by the parametric oscillator itself: Switchings between the $\pi$-shifted oscillating states reduce the overall measured voltage. We performed a separate control measurement that yielded $2.4\pm0.5\,\%$ switching probability, which translates into a maximal fidelity loss of half of that, $1.2\pm0.25\,\%$. The switching rate of the parametric oscillator depends on many parameters, including damping rates and bias points; this error can therefore, with careful engineering, be decreased even further. We could, however, eliminate the effect of phase-switching events by using a rectifying detection scheme, \emph{e.g.}, a diode or a field-programmable gate array (FPGA), tracking the absolute value of the output field instead of its amplitude.\\
\indent The last and smallest contribution to the fidelity loss is the peak separation error, which accounts for the intrinsic overlap between the histograms. However, this contribution is $<\!0.002\%$ for our SNR of 3.39, and can therefore be neglected. For details on the error budget analysis, see Supplementary Note~1 and Fig.~4.\\
\indent By combining the above-mentioned improvements (reduced qubit relaxation rate, optimised qubit manipulations and cooling, enhanced resonator output coupling, and rectifying data acquisition), the read-out fidelity could realistically reach $\approx 99.5\,\%$, limited only by the qubit relaxation.\\
\indent Finally, we demonstrate that  the relaxation time of our qubit is not measurably afflicted by the pump -- see Methods and Supplementary Fig. 5.
Our measurement scheme is, in principle, quantum nondemolition (QND), see Supplementary Note 3;
however, a proper experimental and theoretical assessment of the back-action is outside the scope of this work.\\
\indent Table \ref{tab:OperationSchemes} puts our results in the context of previous work on parametric and nonlinear Josephson amplification and detection circuits.\\
\indent  A flux-pumped, parametric phase-locked oscillator (PPLO) was used as a following amplifier, also enabling sensitive qubit readout~\cite{Lin2014}.
In our work, the qubit was directly coupled to the JPO, which simplifies the experimental setup by reducing the number of microwave components needed.
Also, with a pumping amplitude below the parametric instability threshold, the flux-pumped JPA has been used to read out one qubit~\cite{Lin2013}, as well as multiple qubits coupled to the same bus resonator~\cite{Jeffrey2014}.\\
%
%\textbf{Parametric pumping vs.\@ nonlinear driving.}
\indent There is another way of operating our device: 
instead of pumping the flux at $\omega_p \approx 2\omega_r$, we can apply an alternating pump current ($\epsilon \! = \! 0, \, B(t) \! \neq \! 0$), now at a frequency close to resonance, $\omega_p \approx \omega_r$, and thereby directly modulate the phase difference, $\phi$.
Both methods can provide linear parametric gain upon reflection of a detuned signal ($\omega_s\neq\omega_p/2$ and $\omega_s\neq\omega_p$, respectively).
The flux-pumped JPA has a very wide frequency separation between pump tone and signal, because $\omega_s \approx \omega_r \approx \omega_p/2$, which is a  practical advantage since it makes the resonator's entire instantaneous bandwidth available for amplification with no need to suppress or filter out the pump tone.
Moreover, the $\lambda/4$ resonator has no mode in the vicinity of  $\omega_p$ that the pump might otherwise populate.\\
%
%\indent However, the similarity between the two types of JPA is limited to this case of phase-preserving, linear amplification~\footnote{In phase-preserving amplification, the signal frequency, $\omega_s$, of the input field, $B(t)$, is detuned from the pump. For flux-pumping at $\omega_p \approx 2\omega_r$ (sometimes called degenerate three-wave mixing), $\omega_s = \omega_p/2 + \delta_s$; for current-pumping at $\omega_p \approx \omega_r$ (degenerate four-wave mixing), $\omega_s = \omega_p + \delta_s$. Here,  ``degenerate" refers to the case when both signal and idler ($\omega_i$) frequencies are within one resonant mode, and naturally, $\omega_p = \omega_s + \omega_i$.}.
%
\indent We emphasize that there are indeed two different physical mechanisms in play, since flux and current pumping address orthogonal variables in the sense that
$\Phi =  (\varphi_1 - \varphi_2)\Phi_0/2\pi$ and $\phi = (\varphi_1 + \varphi_2)/2$,
where $\varphi_1$ and $\varphi_2$ denote the gauge-invariant phase differences across the two parallel JJs.
This distinction is also evident in Eq.~(\ref{eq:WS_equation}).
The parametric flux-pumping term, $\epsilon A^*$, modulates the resonant frequency; 
it couples the resonator field amplitude and its complex conjugate, which can provide quadrature squeezing of an input signal and enables phase-sensitive parametric amplification;
and for stronger modulation there is a parametric instability threshold into the JPO regime -- see Fig.~\ref{fig:SetupAndReadoutPrinciple}(b).\\
\indent Current pumping by an input $B(t)$, on the other hand, corresponds to an external force which directly contributes to the intra-resonator field $A$ and drives its nonlinear term $\alpha |A|^2$.
For zero detuning, $\omega_s=\omega_p$, this is the driven Duffing oscillator which has no gain (it offers no phase-sensitive amplification); 
for stronger driving there occurs a dynamical bifurcation but no internal instability or parametric oscillations.\\
\indent Current-pumping with a moderate amplitude is used for linear amplification with the JPA~\cite{Yurke1988,Castellanos-Beltran2008}, which enabled, \emph{e.g.}, the observation of quantum jumps in a qubit~\cite{Vijay2011}.
Current modulation is also used in the latching detection scheme of the Josephson bifurcation amplifier (JBA)~\cite{Siddiqi2004,Siddiqi2005,Lupascu2007,Mallet2009,Schmitt2014}.
There, a higher-amplitude input strongly drives the Duffing nonlinearity near its bifurcation point; the two qubit states can then be mapped onto two different resonator output field amplitudes.
The JBA was used for quantum non-demolition measurement of a qubit, and in a lumped-element resonator \cite{Murch2012}, in which a qubit-state sensitive autoresonance was observed in response to a frequency-chirped current drive.
Yet another method is to couple the qubit to a linear resonator, which inherits a cross-Kerr nonlinearity from the qubit; current pumping of the resonator can then yield a strong output signal that depends on the qubit state~\cite{Reed2010, Boissonneault2010}.\\
\begin{table}[t!]
\begin{ruledtabular}
\begin{tabular}{l | c r r c | c }
Device & $\epsilon$ & $B_s$ & $B_p$ & $\#$ modes & Ref.\\[0.4ex]
\hline \\[-2ex]
JPO$^{(*)}$ & $>\epsilon_{\mbox{\tiny{th}}}$ & 0 & 0 & 1 & This work\\[0.4ex]
JPA & $\lesssim\epsilon_{\mbox{\tiny{th}}}$ & $\neq 0$ & 0 & 1 & \cite{Lin2013}\\[0.4ex]
JPA & $\lesssim\epsilon_{\mbox{\tiny{th}}}$ & $\neq 0$ & 0 & multimode & \cite{Simoen2015}\\[0.4ex]
\hline \\[-2ex]
PPLO & $>\epsilon_{\mbox{\tiny{th}}}$ & $\neq 0$ & $\neq 0$ & 1 & \cite{Lin2014}\\[0.4ex]
\hline \\[-2ex]
JPA & 0 & $\neq 0$  & $\neq 0$ & 1 & \cite{Vijay2011}\\[0.4ex]
JBA$^{(*)}$ & 0 & 0 & $\neq 0$ & 1 & \cite{Mallet2009}\\[0.4ex]
JPC & 0 & $\neq 0$ & $\neq 0$ & 2 & \cite{Bergeal2010}\\[0.4ex]
\end{tabular}
\end{ruledtabular}
\caption{\label{tab:OperationSchemes} Overview of different modes of operation for the various Josephson amplification and detection schemes. The variables  refer to Eq.~(\ref{eq:WS_equation}), where $\epsilon$ denotes the flux-pumping amplitude (at $\omega_p \approx 2\omega_r$), and $B_s$ and $B_p$ denote alternating-current signal and pump amplitudes, respectively (at $\omega_p \approx \omega_r$). The two readout methods marked with an asterisk ($^{*}$) have the qubit directly integrated with the detector, whereas the other devices are used as following amplifiers.}
\end{table}
\indent In conclusion, we have introduced a single-shot readout technique for superconducting qubits $\--$ the Josephson parametric oscillator (JPO) readout.
We demonstrated proof-of-principle operation, obtaining a bare state discrimination of 81.5\,\%.
After correcting for known and reparable errors, this translates into an inferred readout fidelity of $98.7\pm1.2\,\%$, which by implementing a rectifying detection scheme can be further increased by $1.2\pm0.3\,\%$. With foreseeable improvements and optimization, this device would be an attractive candidate for implementing multi-qubit readout in the context of scalable error correction schemes.
This fidelity and the readout time are both amenable to optimization.
\\
\indent Our system integrates a parametric readout mechanism into the resonator to which the qubit is coupled, substantially reducing the number of components needed to perform single-shot readout in a circuit quantum electrodynamics architecture. 
Advantages offered by this readout technique include the potential for multiplexing and scalability with no need for signal-probe inputs, additional microwave circulators, or separate parametric amplifiers. 
As opposed to other integrated readout devices, our pump frequency is far outside of the resonator band and can thus easily be spectrally separated from other transition frequencies in the system.

\section*{Acknowledgements}
The authors would like to thank Jared Cole, G\"oran Johansson, and Baladitya Suri for fruitful discussions. All devices were fabricated in the Nanofabrication Laboratory at MC2, Chalmers. Support came from the Wallenberg foundation, the European Research Council (ERC), the Royal Swedish Academy of Science (KVA), the European project ScaleQIT, STINT, and Marie Curie CIG. The MIT and Lincoln Laboratory portions of this work were sponsored by the Assistant Secretary of Defense for Research \& Engineering under Air Force Contract \#FA8721-05-C-0002. Opinions, interpretations, conclusions and recommendations are those of the author and are not necessarily endorsed by the United States Government.

\section*{Author contributions}
P. K., A. B., M. S., C. M. W., P. D., and J.B. designed the experimental setup. P. K. modeled and fabricated the device. P. K., A. B., S. G., W. D. O., and J. B. carried out the measurements. V. S. gave input on theoretical matters. P. K., P. D., and J. B. wrote the manuscript with input from all co-authors.

\newpage
\section*{Methods}

\textbf{Device fabrication.} We fabricated our device on sapphire, using niobium for the waveguides and the transmon paddles, and shadow-evaporated aluminum for the Josephson junctions. To reduce the surface roughness prior to processing, the 2'' c-plane sapphire wafer was pre-annealed at 1100$^{\circ}$C for 10 h in an atmosphere of N$_2$:O$_2$, 4:1, ramping the temperature by 5$^{\circ}$C/min. The annealed wafer was then sputtered with 80 nm of Nb in a near-UHV magnetron sputter. The first patterning of the sample consists of a photolithography step to define alignment marks and bond pads, deposited using electron beam evaporation of 3 nm Ti and 80 nm Au. Next, the resonator, the transmon islands, and the pump line were defined in the Nb layer using a standard e-beam lithography process at 100 keV, and etched using inductively coupled plasma reactive ion etching (ICP RIE) in NF$_3$-gas.\\
\indent The Al/AlO$_{\small{\mbox{x}}}$/Al Josephson junctions forming the SQUIDs, used to terminate the resonator and for connecting the transmon islands, were then defined in a second e-beam step. After exposure, the 2''-wafer was diced into separate chips, using the exposed e-beam resist as a protective resist. Prior to the first evaporation step, the surfaces of the Nb films where cleaned using in-situ Ar-ion milling inside of the Plassys evaporator. However, due to the substantially different regimes of critical currents, $I_0$, required for the Josephson junction of the transmons and the parametric resonator, two sequential evaporations and oxidations were performed within the same vacuum cycle by rotating a planetary aperture mounted inside the evaporator load-lock, effectively shielding one half of the sample at the time. Finally, a post-deposition ashing step was done to clean the surfaces from organic residues.\\

\textbf{Finding the parametric oscillation threshold.} It is hard to experimentally find the parametric oscillation threshold with good precision, when only considering the parametric oscillation region, Fig. \ref{fig:ParametricRegions}(a), whose observed shape gets smeared by the amplified vacuum noise. In this section we present an alternative method using a weak probe signal: we probe the parametrically amplified response as we sweep the pump amplitude across the instability threshold.\\
\indent We apply a probe signal on resonance, $\omega_s = \omega_{r}^{|0\rangle}$, while applying a detuned pump signal, such that $(\omega_{p} - 2\omega_{s})/2\pi = 100\,$kHz. The signal then undergoes degenerate, phase-preserving parametric amplification (red trace in Supplementary Fig.~1), while the parametric oscillations are cancelled out since we measure the average amplitude of the field. The parametric amplification has maximum gain just at the threshold. We plot the magnitude of the reflected signal as a function of the pump power (at the generator), yielding an oscillation threshold $P_{\mbox{\tiny{th}}} = -10.8\,$dBm, as indicated by the dashed red line. As a comparison, we measure the output power of parametric oscillation (PO), for $\omega_{p} - 2\omega_{r} = 0$ and $B(t) = 0$ --- see the blue trace.\\

\textbf{Limits of the parametric oscillation amplitude.} As briefly discussed in the main text, there are two nonlinear effects that move the resonator away from its pump condition, by means of their associated frequency shifts \cite{Krantz2013},
\begin{equation}
\Delta \omega = - \alpha |A|^2 - \beta \Gamma (\epsilon/\Gamma)^2.
\label{eq:DeltaOmegaShifts}
\end{equation}

The Duffing shift dominates near flux bias $F = \pm \pi/2$; the Duffing parameter is approximated as
\begin{equation}
\alpha(F) \approx \frac{\pi^2 \omega_{\lambda/4}Z_0}{R_K}\left( \frac{\gamma_0}{\cos(F)}\right)^3 = \alpha_0 \left( \frac{\gamma_0}{\cos(F)}\right)^3,
\label{eq:Alpha}
\end{equation}

\noindent where $Z_0 = 50\,\Omega$ is the resonator's characteristic impedance and $R_K = h/e^2$ is the quantum resistance.\\
\indent The pump-induced frequency shift dominates near $F = 0$; it is approximated as
\begin{equation}
\beta(F) \approx \frac{\Gamma}{\omega_{\lambda/4}\gamma_0}\frac{\cos^3(F)}{\sin^2(F)} = \beta_0\frac{\cos^3(F)}{\sin^2(F)}.
\label{eq:Beta}
\end{equation}

The resonator's frequency tuning vs. $F$, Eq.\@ (5) in the main text, is shown in Supplementary Fig. 2(a), for the parameters of our device, and Eqs.\@ (\ref{eq:Alpha}) and (\ref{eq:Beta}) are plotted in Supplementary Fig.\@ 2(b). This figure illustrates that it is essential to bias the system far enough away from the limiting points, $F = 0$ or $\pi/2$, such that neither frequency shift pulls the resonator too far from its pump condition, thereby severely limiting the attainable output power.\\
\indent The steady state solution of Eq. (1) in the main text yields an analytic expression for the expected number of photons within the region of parametric oscillations,
\begin{equation}
\left|A\right|^2 = \frac{\Gamma}{\alpha}\left( \sqrt{\left(\frac{\epsilon}{\Gamma}\right)^2 -1} - \frac{\delta}{\Gamma}\right),
\label{eq:PhotonNumberTheory}
\end{equation}

\noindent which, for our analyzed bias point, amounts to 200$\pm$3 photons in the resonator. From this number, we obtain a Duffing shift $-\alpha |A|^2/2\pi \approx -5.4\pm0.3\,$MHz (for $\alpha/2\pi = 27\pm1.5\,$kHz per photon) and a pump-induced frequency shift $-\beta \Gamma (\epsilon/\Gamma)^2/2\pi \approx -0.64\,$MHz (for $\beta = (7.5\pm0.1)\times10^{-3}$).\\
\indent The parameter $\beta$ has the effect of skewing the parametric oscillation region, yielding an expression for the thresholds plotted in Fig. 1(b),
\begin{equation}
\frac{\epsilon}{\Gamma} = \frac{1}{\sqrt{2}\beta}\sqrt{1 - 2\beta \frac{\delta}{\Gamma} \pm \sqrt{1 - 4\beta \left( \beta + \frac{\delta}{\Gamma}\right)}}.
\label{eq:ParametricInstabilityThreshold}
\end{equation}
\\

\textbf{Calibration of  attenuation and gain via the Duffing non-linearity.} In this section, we present how we calibrated the gain of the amplifier chain, using the photon-number-dependent frequency shift of the Duffing oscillator, $-\alpha |A|^2$, which we recall from the previous section. The frequency of the resonator as a function of input probe power takes the following form,
\begin{equation}
\omega_{r}(P_s) = \omega_{r}(0) - \frac{2 \alpha \Gamma_0}{\Gamma^2}\frac{10^{(P_s-Att-30)/10}}{\hbar \omega_{r}(0)},
\label{eq:DuffingShift}
\end{equation}

\noindent where $\omega_{r}(0)$ denotes the resonant frequency with zero photons in the resonator, $\Gamma_0$ and $\Gamma$ are the external and total loss rates, respectively, and $\alpha$ is the Duffing frequency shift per photon --- recall Eq. (\ref{eq:Alpha}). Using Eq.~(\ref{eq:DuffingShift}), we can fit the extracted resonant frequencies as a function of input probe power at different flux bias points, $F$, with the attenuation, $Att$, as the only fitting parameter (since $\alpha$ can be extracted separately by fitting $\omega_{\lambda/4}$ and $\gamma_0$ --- recall Eq.~(5)). This is shown in Supplementary Fig. 3, where the data for five different flux bias points are fitted to attenuations presented in Supplementary Table 1. From these values, we obtain an average attenuation, $\langle Att \rangle = 127.5\,$dB, which can be compared with the installed 120$\,$dB, indicating that we have a cable loss of 7.5$\,$dB at the measurement frequency. \\
\indent Moreover, from the same measurement we can also obtain an estimate for the gain of the amplifier chain by assuming that all the signal gets reflected when it is far off resonance with the resonator \emph{i.e.} reflection coefficient $|S_{11}|^2 = 1$. Then, the gain is obtained from the relation
\begin{equation}
G = \left| S_{11} \right|^2 + Att.
\label{eq:Gain}
\end{equation}

\noindent For the five gain estimates presented in Supplementary Table 1, we obtain a gain of $G = 81.0\pm0.37\,$dB, at our given bias point. The error bars for this gain estimation has two origins: $\pm0.17\,$dB from the residual of the linear fit to the gain values presented in Supplementary Table 1, and another $\pm0.2\,$dB from the gain drift over time, which can be compared with our 91$\,$dB of installed amplification.\\

\textbf{Calibration of the resonator photon number.} From the obtained calibration of the gain of our amplifier chain, $ G $, we can now calculate the conversion factor between our measured power on the digitizer and the number of photons in the resonator, using the following relation,
\begin{equation}
\left|A\right|^2 = \frac{P_s - P_n}{2 (\Gamma_0/2\pi) \hbar \omega_{r}^{|0\rangle} 10^{G/10}}, 
\label{eq:PowerConversionFactor}
\end{equation}

\noindent where $P_s$ and $P_n$ denote our signal and noise power levels, respectively. We demonstrate this for the bottom panel of Fig.~3, where the resonator is probed at a frequency $\omega_{r}^{|0\rangle}/2\pi = 5.212\,$GHz. The external damping rate is $\Gamma_0/2\pi = 1.02\,$MHz, and we calculate the background power level from the end of the trace (when the pump is off). From the obtained SNR, the number of added noise photons can be estimated accordingly, $\left|A\right|^2/\mbox{SNR}^2 = 16.1\pm1.3$.\\

\textbf{Quantum coherence and readout nondestructiveness.} To study how the parametric pump strength affects the qubit's relaxation time, we here present coherence measurements for the transmon. First, we calibrate a qubit pulse duration corresponding to a $\pi$-pulse, using a Rabi measurement, where the pulse duration time is swept, for a fixed pulse amplitude. From the fit in Supplementary Fig.~5(a), a pulse length of $\tau_{\pi} = 52\,$ns was obtained, and the Rabi decay time was $T_{\mbox{\tiny{rabi}}} = 2.53\pm0.15\,\mu$s. The histograms corresponding to the first 0.5$\,\mu$s are plotted in Supplementary Fig. 5(b), using the same projective technique as for the histograms in Fig. 5(c) in the main text. Finally, we perform a set of $T_1$ measurements for different pump amplitudes $\epsilon/\Gamma$, and compare these with traditional reflection readout, where we apply a weak resonant probe signal, but no pump ($B(t)\neq 0, \epsilon=0$). The fits to the relaxation times suggest that our readout is not any more destructive to the quantum state of the transmon than the traditional readout technique is. We note, however, that our extracted relaxation time is limited by the Purcell effect, yielding $T_1 \approx [2 \Gamma_0(g_{01}/\Delta)^2]^{-1} = 4.11\,\mu$s. Also see Supplementary Note 3.\\

\newpage
%\bibliography{References}

\end{document}